\documentclass{IEEEtran}
\usepackage{cite}
\usepackage{amsmath,amssymb,amsfonts}
\usepackage{algorithmic}
\usepackage{url}
\usepackage{graphicx}
\usepackage{textcomp}
\def\BibTeX{{\rm B\kern-.05em{\sc i\kern-.025em b}\kern-.08em
    T\kern-.1667em\lower.7ex\hbox{E}\kern-.125emX}}
\begin{document}

\title{6G Communications: A Vision on the Potential Applications}

\author{Sabuzima Nayak and Ripon Patgiri, \IEEEmembership{Senior~Member,~IEEE} \\National Institute of Technology Silchar

\thanks{Sabuzima Nayak and Ripon Patgiri, Department of Computer Science \& Engineering, National Institute of Technology Silchar, Cachar-788010, Assam, India, Email: \textit{sabuzimanayak@gmail.com} and \textit{ripon@cse.nits.ac.in} }
}

\maketitle

\begin{abstract}
6G communication technology is a revolutionary technology that will revolutionize many technologies and applications. Furthermore, it will be truly AI-driven and will carry on intelligent space. Hence, it will enable Internet of Everything (IoE) which will also impact many technologies and applications. 6G communication technology promises high Quality of Services (QoS) and high Quality of Experiences (QoE). With the combination of IoE and 6G communication technology, number of applications will be exploded in the coming future, particularly, vehicles, drones, homes, cities, hospitals, and so on, and there will be no untouched area. Thence, it is expected that many existing technologies will fully depend on 6G communication technology and enhance their performances. 6G communication technology will prove as game changer communication technology in many fields and will be capable to influence many applications. Therefore, we envision the potential applications of 6G communication technology in the near future.
\end{abstract}


%%%%%%%%%%%%%%%%%%%%%%%%%%%%%%%%%%%%%
%% IEEE Communication Magazine, IF 10.356
%%%%%%%%%%%%%%%%%%%%%%%%%%%%%%%%%%%%%
%% Word limit: 5500 includes everything and 7 pages
%% Reference limit: 15
%Intro approx: 700 words
%abstract:100 words
%rest: 4700 OK
%%%%%%%%%%%%%%%%%%%%%%%%%%%%%%%%%%%%%

\begin{figure*}[!ht]
    \centering
    \includegraphics[width=\textwidth]{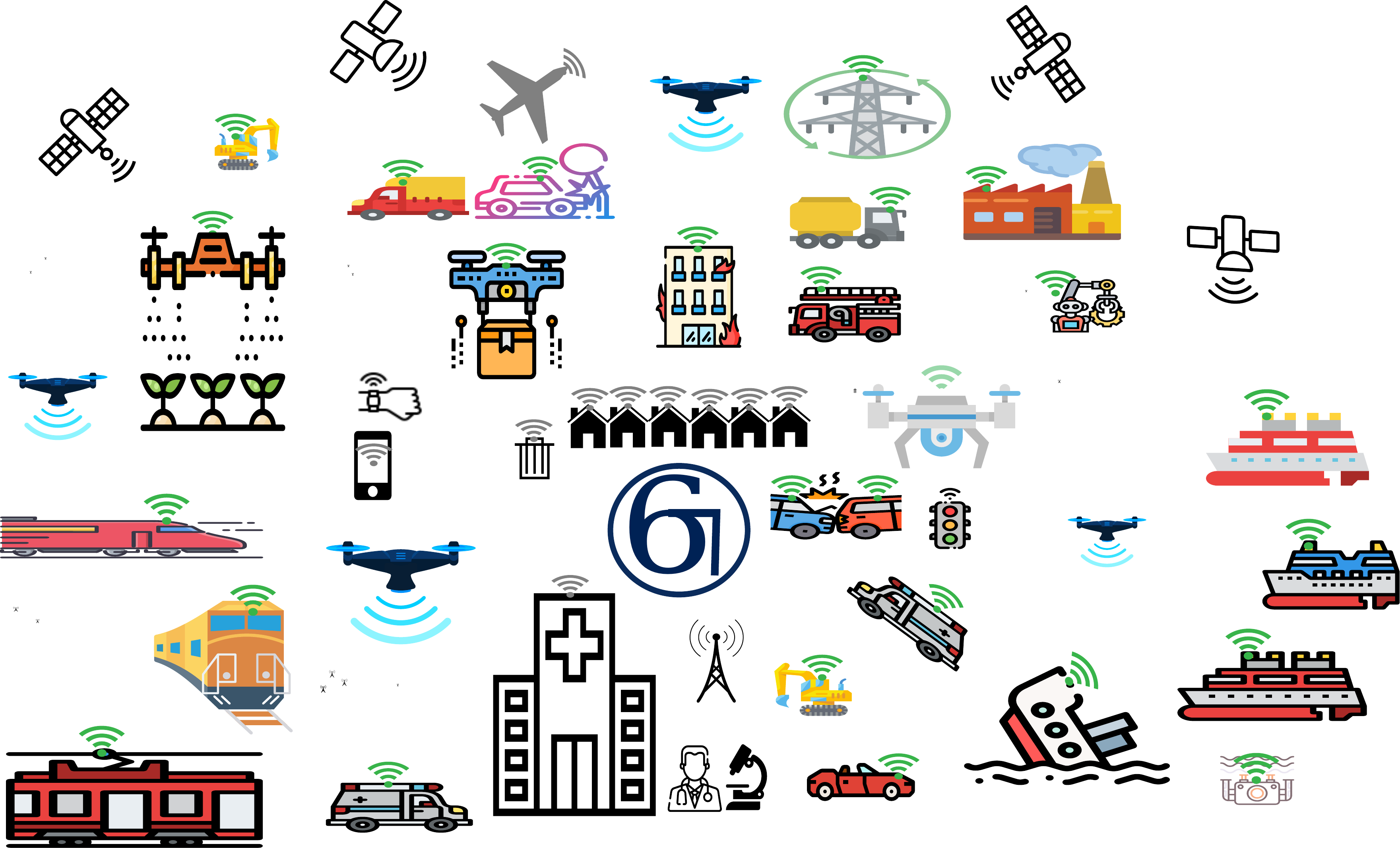}
    \caption{6G communication technology enables Internet of Everything (IoE). The landscape of Applications of IoE using 6G communication technology. It depicts future dependability of intelligent vehicles, drones, transportation, hospital, agriculture, trains, ships, and industry on 6G communication technology. 6G will not only a smart phone communication anymore, but also a default communication system for many devices and technologies.}
    \label{fig1}
\end{figure*}

% HERE, WE HAVE TO EMPHASIZE ON ``HOW 6G COMMUNICATION TECHNOLOGY IMPACT IN THE APPLICATIONS FIELD AND REVOLUTIONIZE''
6G mobile communication technology is one of the most prominent emerging research area and radical technology, which will change our perception on lifestyle, society and business. With the advent of communication technology, 5G is maturing, and thusly, it demands next generation mobile communication technology. Since, we have already evidence that every decade there is a mobile generation, so, it is expected that 5G will serve from 2020 to 2030, 6G will serve from 2030 to 2040, and 7G will serve from 2040 to 2050. Currently, 5G is deployed in few countries. 5G communication technology is also able impact on the global economy. It is anticipated that the next generation of 5G will be able to create more impact on the global economy. Withal, the requirements of 6G are extremely different from 5G communications \cite{Chettri}. Consequently, the parameters of 6G will enable tremendous new applications and technologies \cite{Saad} as depicted in Figure \ref{fig1}. For instance, 6G will be a prominent player in vehicular technology, healthcare and commercial enterprise. Also, 6G communication technology will be able to influence many researchers, practitioners and industrialists. Hence, there are diverse research articles already discussed on parameters of 6G \cite{Katz,Giordani,Gui,Chen,Giordani}. Moreover, Q. Bi highlighted ten trends of 6G communications \cite{Bi}. Kato \textit{et. al.} exposes ten great challenges on integration AI with 6G communication technology. Letaief \textit{et. al.} envisioned the truly AI-driven 6G communication technology \cite{Letaief}. 

The promises of 6G communication technology create numerous issues and challenges \cite{Dang,Zong}, albeit, worldwide deployment is expected from 2030. Moreover, rural deployment of such technology is a grand challenge \cite{Yaacoub}. Currently, 5G communication technology focuses on campus solution, and thereby, the mobility and coverage are issues. Therefore, 5G is unable to support many applications. Nonetheless, the 6G communication technology promises increase coverage and mobility using satellite communication. Presently, we are lacking $\geq 1$ Tbps data rates and extremely reliable and low latency, and this triggers compromising in QoS. As per the requirements and promises of 6G communications, there are numerous research possibilities and opportunities which will endure many new applications to be conceived. Thus, 6G will bloom the global economy, society and lives.

In the 6G communication era, many existing technologies will be redefined and restructured. Reshaping existing technology will shift our lifestyle, society and business. Prominently, Internet of Things (IoT) will be redefined as Internet of Everything (IoE) which will be the dawn of many novel technologies. IoE will be a key player in the 6G communication era, and it will enable intelligent city, intelligent healthcare (Intelligent Wearable devices and Intelligent Internet of Medical Things), intelligent Industrial Internet of Everything (IIIoE), intelligent grid, and Intelligent Robots. We also foresee that the 6G communication technology will multiply in spawning many new applications. Therefore, in this article, we envision the impact of 6G communication technology in various fields. Thus, Section \ref{6G} briefs on 6G technology to highlight the key features. Besides, Section \ref{ena} explores the enabling technology of 6G communications. Sections \ref{veh}, \ref{rob}, \ref{vir}, \ref{hea} and \ref{cit} exposes the impact of 6G communications technology on Vehicular Technology, Robotics, Virtual-reality, Healthcare and City. Likewise, Section \ref{ind} exposes on how 6G communication technology will be able to enable an Industrial revolution. Furthermore, Sections \ref{nan} and \ref{dat} exposes that the nano-things and datacenters depends on the 6G communication technology.  Finally, Section \ref{con} draws a suitable conclusion.

%%%%%%%%%%%%%%%%%%%%%%%%%%%%%%%%%%%%%%%%%%%%%%
%% 6G Technology
%%%%%%%%%%%%%%%%%%%%%%%%%%%%%%%%%%%%%%%%%%%%%%
\section{6G Technology}
\label{6G}
6G communication technology has a potential to change the definition of our perception of life, lifestyle, society and business. 6G communication is all about sixth sense communication \cite{Kantola}. Therefore, many countries have already begun a project on 6G communication technology, particularly, Finland, USA, South Korea, China \cite{Dang} and Japan \cite{Docomo}. Its prerequisites are very challenging and numerous requirements have been already discussed through various articles \cite{Katz,Katz1,Nayak,David,Yang}. Particularly, 6G communication technology will operate at (THz) frequency to gain 1 Tbps data rate \cite{Kato,Elmeadawy,Rappaport,Tomkos}. However, 1 Tbps data rate is not enough for holographic communications. It demands more data rate to provide full holographic communication support. The wavelength frequency is expected to be $300 \mu m$ to $100\mu m$. Researchers are exploring to increase the efficiency of THz signal by spectrum reusing and sharing. Some techniques already exist for spectrum reuse such as cognitive radio (CR). It helps many wireless systems to access the same spectrum through a spectrum sensing and interference management mechanisms \cite{Chen}. In case of spectrum sharing, temporally underutilized or unlicensed spectrum is utilized to maintain availability and reliability. Symbiotic radio (SR) is a new technique to support intelligent, heterogeneous wireless networks. It will aid in  efficient spectrum sharing. Even so, deploying these techniques in the 6G wireless network are still big challenges.  Also, 6G is expected to deliver truly AI-driven communication technology \cite{Nayak,Kato,Nawaz,Piran}. In addition, 6G promises to provide high security, secrecy and privacy \cite{Dang}.  Likewise, it will be three dimensional communication technology, particularly, time, space and frequency. 

\subsection{Transition from Smart to Intelligent} 
The IoE will be intelligent, thereby, 6G will also be intelligent. Consequently, all smart devices will be converted to intelligent devices and these intelligent devices will be truly AI-driven devices \cite{Nayak1}. Thusly, the intelligent device (may be tiny device) will be able to predict, make a decision and share their experience with other intelligent devices. So, there is a paradigm shift from smart to intelligent era using 6G communication technology and AI. 

\subsection{Quality of Services}
6G Technology promises to provide high Quality of Services (QoS) and the parameters includes high data rate of at least 1 Tbps, extremely reliable, further-enhanced mobile broadband (FeMBB), low latency communication (ERLLC), long-distance and high-mobility communications (LDHMC) ultra-massive machine-type communications (umMTC) and  extremely low-power communications (ELPC) \cite{Zhang1}. Moreover, QoS includes massive broad bandwidth machine type (mBBMT), mobile broad bandwidth and low latency (MBBLL),  and massive low latency machine type (mLLMT) \cite{Gui}. These parameters will serve diverse applications to revolutionize. Thus, the 6G technology will be a revolutionary technology in many domains. In addition, high QoS will enable many new applications, for instance, telesurgery. Also, 6G technology promises high QoE along with high QoS which defines user-centric communications. It will be achieved by holographic communications, augmented and virtual reality, five sense communications and tactile Internet. Moreover, QoE will bring revolution in intelligent devices, intelligent cars, intelligent drones, intelligent ambulances, and many more \cite{Nayak}. Achieving high QoE depends on the implementation of all desired parameters by 6G technology. Similarly,  Quality of Life (QoL) is determined to enhance the lifestyle with QoS and QoE in healthcare. 6G technology will enable the high QoL using communication technology. The key parameters of QoL are intelligent ambulance services, remote health monitoring of patients, including elderly persons, intelligent accident detection, Hospital-to-Home (H2H) service, telesurgery and precision medicine. 

%%%%%%%%%%%%%%%%%%%%%%%%%%%%%%%%%%%%%%%%%%%%%%
%% 6G Enabling Technology
%%%%%%%%%%%%%%%%%%%%%%%%%%%%%%%%%%%%%%%%%%%%%%
\section{6G Enabling Technology}
\label{ena}
\subsection{Internet of Everything (IoE)} %Done 
6G communication follows 6C’s, namely, capture, communicate, cache, cognition, compute, and control \cite{Gui}. Cognition helps in formulating feasible determinations based on input digital data. These are intelligent determinations that make the computing easy. Then, computed data are transmitted to intelligent devices to control the action taken by the devices \cite{Gui}. For example, raising an alarm. IoE will use the core service of combined and enhanced eMBB and mMTC.  Requirement of IoE from 6G are huge capacity to connect millions of intelligent devices and high data rates to support touch experiences in those devices. Industrial IoT (IIoT) uses the core service of combined and enhanced URLLC and mMTC. Huge capacity to connect the sensors, actuators and staff for communication. And, low latency to maintain seamless integration among them \cite{Gui}. When 6G will be commercial, it will be the era of Big data 2.0 that requires a supercomputer to compute and analyze data \cite{Nayak1}.

\subsection{Edge Intelligence}
6G will rely on Edge computing \cite{Shi1} to bring the Cloud features closer to intelligent devices. Edge technology will offer uninterrupted and high speed Internet services to the intelligent devices. It collects, computes and analyses the data in real time in Edge nodes. Edge node also filters data and transmits only important information to the Cloud for storage. Edge technology reduces communication and computation cost. Some other advantages are low latency, reliability, adaptability, scalability and privacy. All the advantages of Edge technology will greatly help 6G to meet its prerequisites to provide high QoS. Edge Intelligence is combining Edge computing and AI \cite{Zhou}. Edge analytics, implement AI algorithms for analysis in Edge nodes \cite{Shi}. With the help of AI, Edge nodes are capable of developing image, data and video Edge analytics. Execution of AI algorithms requires high computational resources, and power consumption, which is limited in the Edge nodes. One of the network nodes in 6G will be Edge nodes. Moreover, 6G will have real-time intelligent Edge that computes and analyze on live data \cite{Huang}.

\subsection{Artificial Intelligence}
6G will be a truly AI-driven communication network, causing the system self-aware, self-compute and self-decide on a situation \cite{Nayak,Elsayed}. 6G will provide global coverage, i.e., space-air-water. 6G will achieve this goal by making different aspects of communication ``intelligent”, and, intelligence means AI. At the physical layer of the network, AI will help channel state estimation and prediction, automatic modulation classification, adaptive encoding and decoding, and intelligent beamforming. Likewise, physical layer security \cite{Hamamreh} is key important features of 6G communication technology. At the data link layer, resource allocation that implements deep reinforcement learning \cite{Kato}. Likewise, at the transport layer, route computing and intelligent traffic prediction algorithms will be deployed \cite{Gui}. 6G will explore dynamic spectrum access. 6G will depend on AI for bringing the heavy computation responsibility to provide next generation services.

%%%%%%%%%%%%%%%%%%%%%%%%%%%%%%%%%%%%%%%%%%%%%%
%% Vehicular Technology
%%%%%%%%%%%%%%%%%%%%%%%%%%%%%%%%%%%%%%%%%%%%%%
\section{Vehicular Technology} 
\label{veh}
Many vehicles can be driverless and intelligent \cite{Viswanathan}. The intelligent vehicles will optimize fuel consumption, route and work efficiency, because 6G communication technology can provide real-time services. Intelligent vehicles are more economical, because they can predict future and optimize the problem economically. The future vehicle will be featured with truly AI-driven and high speed wireless communications. The future vehicles will also be integrated with diverse sensors. The vehicle will be intelligent with the help of sensor devices and communication technology. 

\subsection{Intelligent Cars} 
It is expected that 6G will be able to deliver self-driven vehicle, i.e., steering-less vehicles. The vehicles will be truly AI-driven learning from practical experiences \cite{Tang}. The cars will monitor passenger health, for example, blood pressures, heartbeat rate, body temperature, brain wave capturing, and emotion detection due to truly real-time communication capability of 6G. Also, the intelligent cars will turn into an intelligent ambulance in case the passenger need treatment during the ride. It is achieved by consulting the remote doctors using mix-virtual reality/Holograph through 6G communication. Intelligent cars will avoid accidents by exchanging information with nearby vehicles. Batteries will be charged without the wire in a mobile state by UAVs \cite{Zhang}. Achieving a complete intelligent driving is really difficult because its prerequisite is a combination of many complex algorithms. Instances of such complex algorithms are automatic driving, path planning, obstacle detection, and vehicle monitoring \cite{Gui}. It will require the support of high data rate, low latency, seamless connectivity and integration of AI. It will produce a huge amount of data and also upload these on the Internet in real-time.

\subsection{Unmanned Aerial Vehicles}
UAV will be used for many purposes \cite{Li}, for instance, fronthaul and backhaul services in rural area. The intelligent Drones will be capable of communication in Drone-to-Drone (D2D) and Drone-to-Infrastructures (D2I). Intelligent Drones can share their knowledge. Also, it will be engaged for faster deliver of online products. Moreover, it will be utilized in critical operations where human cannot reach. Police will use Drones to catch a thief. Also, the surveillance intelligent Drone will continuously monitor the populated location. These types of intelligent Drones will detect fire, accidents, traffics, etc., in real-time so that police will be informed in real-time. Police can also employ the intelligent Drones to spray the tear gas to control the mobs. Also, drones can be useful in agriculture.

Currently, Closed Circuit Televisions (CCTV) are used for security and surveillance which makes us very secure. Moreover, the CCTV is also utilized as a witness in court in many countries. However, CCTVs are not efficacious in securing many things and identifying the threats. For instance, CCTV is installed in a fixed pole or wall and they cannot chase any thief. Therefore, the Drone will be added to fortify the security and surveillance systems. For illustration, a thief or terrorist can be chased. The Drone can stream the videos continuously to the monitoring rooms. This requires high speed data link to provide high definition videos in real-time. Also, intelligent Drone can alert possible threat to the security force. The Drone will play a prominent role not only in the city, but also in border areas. It is very dangerous in Line of Control (LOC) to monitor the enemies for 24$\times$7 mode. The Drone can provide the service of security and surveillance in extremely dangerous border locations. Thus, Drone can enhance border security dramatically in defense sector such dangerous areas. It requires 6G communication technology surveillance, however, 5G communication technology can also solve, but QoS will be lower.

\subsection{Intelligent Transportation} 
Achieving a complete intelligent driving is very difficult because its prerequisite is a combination of many complex algorithms. For example, automatic driving, path planning, obstacle detection, vehicle monitoring and emergency rescue operations. It requires high data rate, low latency, seamless connectivity, etc., and it can be provided by 6G communication technology. It will produce a huge amount of data and also upload these data on the Internet in real-time.

\subsubsection{Domestic transportation} 
Intelligent domestic transportation will remain connected to the 6G Internet during the whole journey. It will even provide real time information. 6G with its high speed internet, low latency and global coverage will provide services smoothly without interruption. Therefore, reliable and real time information will be provided to the passenger and reducing their waiting time. The real time data will also help the driver to remain updated about the traffic or any blockage in the route. In case of heavy traffic it will help drivers to drive safely. And, blockage in the route due to an accident or natural disaster will be known earlier. Based on the information the driver can reroute the path and reduce harmful gas emission. Another key point is maintaining security. Police can use the camera present in public transportation to track a suspected vehicle. In case of small public transportation, eg., cab, taxi the passenger can detect the route taken by the driver. Moreover, in case of any threat the passenger can inform the police without any direct call. The intelligent vehicles also continuously monitor its own components and parameters. Upon detection of any damage or abnormality, then immediate action is taken.
%%%%Gingered%%%% Not seen missed it.

%%% To be done!  
\subsubsection{International Transportation}
6G will help in keeping track of vehicles crossing the border of a country by the security personnel. This will help in reducing crimes such as human trafficking and smuggling. 6G will provide Internet services during airplane and ship journey. Airplanes and ships will also remain connected to 6G IoT. The 6G provides global coverage, thus, predicting any sudden changes in weather such as rain or thundering. This real time information will help to take necessary actions. With intelligent transportation the airplane and ships can be made fully automated. Moreover, global coverage will help tracking the airplane and ships in case of any accident. And, immediate medical and rescue services can be provided. In addition, 6G will provide services in underwater. Therefore, a ship or airplane can be tracked in case it has drowned and got displaced due to strong water current. 

%%%%%%%%%%%%%%%%%%%%%%%%%%%%%%%%%%%%%%%%%%%%%%
%% Robotic communication
%%%%%%%%%%%%%%%%%%%%%%%%%%%%%%%%%%%%%%%%%%%%%%
\section{Intelligent Robotic communication} 
\label{rob}
Robotic communication refers to communication between robots or between humans and robots. Ad-hoc networks are always used for robotic communication because robots are independent nodes. Robots depend on the AI for communication. Robots are greatly helpful in home services. It is used as personal assistance for physically challenged people and elderly services. Intelligent Robot can learn to serve efficiently as personal assistance. To achieve learning Robot, AI and Machine Learning has to be integrated. Robots can be a great teacher as well as a learner, which will require mobile communication connection to the Internet. Intelligent Robots can share their experience with other intelligent Robots. Therefore, Robot-to-Robot (R2R) communications can improve intelligent robots in service. Moreover, intelligent Robots can be used in the industry for economical, efficient and faster manufacturing of products.

\subsection{Aerospace Robotic communication} 
There is also a plan to develop an intelligence space. An intelligence space is an intelligence environment to monitor and control activities in space and smooth communicate with people. The maintenance will mostly depend on robots \cite{alsamhi}. Robots will observe the environment and compute the data to take action in case of extreme situations. They also have to constantly transmit data to Earth. Therefore, to achieve such communication 6G will be required to have a smooth and high speed communication. Moreover, in the intelligence space, the robots will be distributed. Thus, they require a reliable and dynamic communication network. Robots will also transmit images and videos. Transmitting such high sized data compared to files requires high speed and high data rate of 6G. Moreover, the features of 6G such as high data rate, speed, high QoS will help to provide a 3D image of the space for a better understanding of the space activities. 

\subsection{Underwater Robotic communication}
Robotics is also explored for underwater tasks such as search and rescue, imaging and security.  The terrorists also use waterways for entering the country. So, constant surveillance is essential. In such cases sometimes, robots will be more efficient. However, the underwater robots have to constantly interact with the ground station. 6G will be very helpful for such communication. A smooth and reliable communication channel needs to be established for real time analysis and action planning. 6G high speed and low latency will help in obtaining images and video data. It will also help in constantly locating the robot on the map and providing navigation instructions. 

%%%%%%%%%%%%%%%%%%%%%%%%%%%%%%%%%%%%%%%%%%%%%%
%% Virtual communication
%%%%%%%%%%%%%%%%%%%%%%%%%%%%%%%%%%%%%%%%%%%%%%
\section{Virtual communication}
\label{vir}
The QoS provided by 6G communication technology will satisfy many requirements of virtual communication such as holographic communication, augmented reality and virtual reality. It requires uninterrupted high speed, low latency to constantly maintain virtual presence, touch, experience etc. 

\subsection{Holographic communication}
Holographic communication requires high data rates to provide good QoS and also streaming videos of high definition. It also requires very low latency for real-time voices and spontaneous control responses \cite{Giordani,Strinati}. 6G will be capable of satisfying all the requirements. Holographic communication will use the core service of combined and enhanced eMBB and URLLC. Holographic communication will be a major breakthrough during 6G. Due to high speed and uninterrupted communication, communicating with a person through holographic communication will be similar to face-to-face interaction. The person will be able to move around without anyone’s intervention. Physical presence at a formal meeting, or interview will not be compulsory. 

\subsection{Augmented Reality and Virtual Reality}
The requirements of augmented reality (AR) and virtual reality (VR) is similar to holographic communication. The core service used by AR and VR is combined and enhanced eMBB and URLLC. High data rates to provide good QoS and high definition videos. Very low latency is essential for real-time voices and immediate control responses \cite{Gui}. Peak data rate requirement of AR and VR is 1 Tbps and user experience of $>$10 Gbps with $>$0.1 ms latency which can be provided by MBBLL \cite{Gui}. AR provides visibility inside the object without separating the different components of the object. For example, various components of a machine are visible without removing the machine parts. Moreover, the depth of the visibility can be changed, i.e. user can view deeper and deeper inside the object (physical depth). Using AR the viewed part is also zoomed for better visibility. VR helps in creating one’s own artificial environment. It helps in adding details related to games, movies, or animation. Moreover, it will also help in better understanding of the documents. Moreover, due to 6G the holographic communication, AR and VR will be combined for better collaborative works. 

\subsection{Tactile/Haptic Internet}
Tactile Internet requires ERLLC and high speed communication to grab the tactile in real-time. Usually the remote user sends the signal to another human or robot. This technology is applicable where a physical presence is compulsory to perform a task. Usually this task requires precise work and robots cannot always be trained to perform such tasks. For example, surgery or defusing the bomb. During telegery, the doctor can use tactile technology to perform a surgery using a human or a robot. As per the movement of the remote doctor the robot will move. Such detailed work required uninterrupted and high speed Internet. 6G will provide great support in achieving these goals.

%%%%%%%%%%%%%%%%%%%%%%%%%%%%%%%%%%%%%%%%%%%%%%
%% Healthcare
%%%%%%%%%%%%%%%%%%%%%%%%%%%%%%%%%%%%%%%%%%%%%%
\section{Intelligent Healthcare}
\label{hea}
QoL will be the greatest player in ensuring intelligent healthcare. It is expected that 6G communication will revolutionize healthcare. 6G will be able to overcome the time and space barriers of healthcare in the near future. Intelligent healthcare will implements Hospital-to-Home (H2H) service which is implemented upon intelligent vehicle. This mobile hospital will replace ambulance services in the near future. H2H will also implement real-time accident detection and automatic emergency detection. Diverse intelligent wearable device and sensors will help in the detection of accident automatically in real-time. Also, needle-free Blood Sample Reader (BSR) sensor will greatly help in the detection of diseases automatically \cite{Nayak2}. BSR sensor will read the blood sample and send it to the pathology laboratory. Therefore, Intelligent Wearable Devices (IWD) reduce the risk of medical staff in contacting with viruses. Also, the patient is not required to travel to the hospital for blood tests.

\subsection{Intelligent Internet of medical Things (IIoMT)}
Moreover, Intelligent Internet of Medical Things (IIoMT) will help in avoiding the time and space barriers \cite{Nayak2}. For instance, remote doctor can perform surgical operations using telesurgery which requires high speed communication. The doctor can supervise the telesurgery through verbal, telestration or tele-assist \cite{HUNG}. For verbal, doctor(s) will use holographic communication. Hologram helps the doctor in the movement to have a better visual of the surgery area. For telestration, doctor(s) will use AR and VR. And, the doctor(s) can tele-assist the surgery using tactile/haptic technology. The requirements of telesurgery can be fulfilled by 6G communication technology and 6G will prove that surgery can be performed beyond boundaries.

%%%%%%%%%%%%%%%%%%%%%%%%%%%%%%%%%%%%%%%%%%%%%%
%% Intelligent City
%%%%%%%%%%%%%%%%%%%%%%%%%%%%%%%%%%%%%%%%%%%%%%
\section{Intelligent City}
\label{cit}
The key requirement of intelligent city is a real-time communication system. Intelligent City manages intelligently in real-time the traffic, waste, home, grid and environment which requires high QoS from wireless communication technology.

\subsection{Intelligent Traffic} 
Smart cities, mainly focus on enhancing theQoL. Intelligent cities will also focus on the environment. Therefore, intelligent traffic is required because longer the vehicle will halt in traffic more harmful gases will be released \cite{Cikhardtova}. Hence, in future the traffic signals will also be connected to IoE.The Intelligent traffic signal will analyze the data to determine the average traffic opening duration for smooth movement of vehicles. This duration will be different for different roads based on real-time data. In case an ambulance is traveling for an emergency then it can send notification to intelligent traffic signals. Upon receiving the notification the intelligent traffic signal will track the ambulance using GPS. At appropriate times the traffic will open for uninterrupted ambulance movement. For such applications, high speed and low latency 6G Internet will be essential. Similarly, the police or army vehicles will be provided an uninterrupted movement. Moreover, intelligent traffic signals can also be used to interrupt any suspected vehicle. They will also monitor the vehicles for any traffic rule violation. In addition, whenever any dispute happens among people the intelligent traffic signal can immediately alert the police. Moreover, camera present in traffic signals can monitor for any accident that happens in an isolated and remote area. In such situations, the live data have to be analyzed quickly for informing appropriate authority. And, 6G Edge nodes will be capable of performing those tasks. 

\subsection{Intelligent Waste Management}
Intelligent waste management is also a part of an intelligent city to reduce pollution and health risk. The dustbins will have sensors to determine whether it is full. In case it is full, the waste collection vehicle will be notified. It will collect the waste quickly. In some areas the dustbins may get full quickly, hence, it requires waste pickup twice or more in a day. And, in some localities the dustbin may be getting full after two days or more. Therefore, intelligent waste collection will help in optimizing utilization of resources. All the dustbin sensors will be connected to IoE. 6G global coverage will help to connect every corner of the city. 

\subsection{Intelligent Home} 
An intelligent home assistant will connect all intelligent home appliances such as television, home security, washing machine, music system to IoE. Human presence and some commands will activate the appliances and allow the people of the house to remotely access the appliance. An intelligent light will switch off in the absence of any person in the room or maintain the light intensity based on sunlight intensity. This is very useful for small children and elderly people. During an emergency, an intelligent home assistant will detect the issue and inform the appropriate authority. For example, a fire at a dangerous level the water sprinkler is automatically opened or informs the fire station in case fire is not under control. The intelligent home will also have an intelligent environment monitoring system. It will keep track of the environmental condition of the home. For example, if the weather is hot then the central air conditioner will start to keep the home cool. This helps in keeping the people of the house healthy. If a family member is ill, then that room's environmental condition will be different from the rest of the rooms. A home will have many intelligent home appliances, all these will be connected to IoE. The data generated will be huge and these need to be processed by 6G Edge nodes quickly. The people of the home can use intelligent phones or some wearable to physically control the intelligent home appliances. During security related situations such as any forced entry or fire, 6G Edge nodes have to compute the data and inform the appropriate authority quickly even without the permission of the owner. 

\subsection{Intelligent Grid}
Intelligent grid is important because 6G nodes will perform heavy computation and will require periodic or constant energy supply. The most important point is to provide global coverage, 6G will have high density, therefore, an intelligent grid is very essential for the 6G communication system. The intelligent grid is AI based, fully automated, remote control and has self-healing features. With the exhaustion of natural resources, intelligent grid will rely completely on renewable energy resources. It will intelligently integrate the distributed and renewable energy resources. 6G nodes will both depend and help in the power grid. The intelligent 6G nodes placed at different locations can determine the power consumption during different times of the year or season. These data will be used to determine the power production requirement for an intelligent city. After sunset the 6G nodes will automatically ON the street light. One important point is that an intelligent home will try to reduce electricity consumption, thus, optimizing the power supply. Intelligent grid will also store energy in case of emergencies. 6G nodes will intelligently switch to reserve energy in case of power cuts. Another concept is micro grid. Using renewable energy resources, the consumers will also produce electricity. Such consumers are called prosumers. Prosumers will sell excess electricity to the grid. Such a small grid is called a micro grid \cite{Atasoy}. Moreover, in case of a power cut the intelligent nodes will take the electricity from the micro grid. With 6G connectivity, the monitoring and maintenance of the micro grid are easy. Real time electricity reserve in the micro grid is recorded and saved. Any illegal sale of electricity, such as industries or illegal institutions doing illegal activity is monitored. The electricity billing will also be intelligent, the intelligent home will send the electricity consumption details to the power grid datacenters and a bill will be automatically forwarded to the consumer. 

%%%%%%%%%%%%%%%%%%%%%%%%%%%%%%%%%%%%%%%%%%%%%%
%% Industrial revolution
%%%%%%%%%%%%%%%%%%%%%%%%%%%%%%%%%%%%%%%%%%%%%%
\section{Industrial revolution}
\label{ind}
The Industry 4.0 was revolutionized as digitalization \cite{Ben}. Industry 5.0 is about personalization and human-centric manufacturing. Industry 5.0 will be able to integrate AI and 5G communication to increase productivity. Moreover, the industrial Robots will become more smart due to the advent of AI and communication technology. Ultra-massive Machine-Type communications take place in the industry due to the huge number of sensors and robots. The robots and sensors require extremely reliable and low latency communication for precision. Also, communications help in faster production. Therefore, it is expected that productivity will increase in Industry 5.0. Moreover, intelligent Robots can optimize their works efficiently and economically. The Industry focuses on all the processes from procuring the raw materials to delivery of finished product to customers. Therefore, an industry has to solve many complex problems to reduce overall profit. The communication system has to support industrial IoT (IIoT). IIoT application focuses on power system monitoring, demand-side energy management, integration of renewable energy generators and coordination of distributed power storage \cite{Hasan}. Moreover, the sensors and robots will produce huge volumes of data. These data will be processed in Edge nodes \cite{Wu}. However, Industry 5.0 will not be capable of supporting Industries completely due to the presence of many issues. But, it is expected that Industry 6.0 will be about truly AI-driven industry. 6G Edge nodes will be able to handle heavy computation, hence, providing spontaneous responses. 6G with high density will be able to manage a massive number of robots and sensors. Industry 6.0 will deliver Industrial Internet of Everything (IIoE) \cite{Nayak}. Thus, Industry 6.0 will be the Intelligentization of Industry.

\section{Internet of Nano-Things} 
\label{nan}
Nano-things have components of nano size, hence, they are capable of performing simple tasks. It has small memory, thus, data storage capacity is low. A communication network consisting of nano-things is called the Internet of Nano-Things (IoNT) \cite{Akyildiz1}. Nano-things communicate within short range. Therefore, within a small range more nano-nodes need to be deployed. Communication of IoNT can be performed by using molecular or THz communication. However, THz communication will be secure, reliable and faster compared to molecular communication \cite{Sicari}. For THz communication, 6G will be an excellent choice. Therefore, 6G will support IoNT and will make it possible for deployment. Nano-things are small and simple, hence, speed will be more. In this regard, 6G will $>$1Tbps speed will perfectly complement IoNT. 6G technology will have high density. Hence, controlling the IoNT network with a massive number of nano-things will be easy. Another important point is in some situations the nano-things have to continuously generate valuable information, but due to small memory these data need to be transmitted to the datacenter. Therefore, with high speed 6G communication network nano-things will transmit the data smoothly.

%%%%%%%%%%%%%%%%%%%%%%%%%%%%%%%%%%%%%%%%%%%%%%%%%%%%%%%%
\section{Datacenter connectivity} 
\label{dat}
6G with its high QoS will provide great support for the development of many technologies. For instance, holographic communication, IoNT, telesurgery. 6G will deploy a massive number of network nodes such as Edge nodes, satellites, drones, etc. to provide high density and global coverage. These 6G network nodes will be capable of computation and analysis. However, these nodes have smaller power supplies and memory. The data generated by the 6G network nodes are high in both volume and quality. For example, underwater rescue operations stream a real time high quality video. This high sized data is not possible to store in the 6G network nodes. Thus, the data are transmitted to the datacenter for storage. Datacenter requires high data rate and extremely low latency \cite{Rommel}. Due to the high speed of 6G the data generation and data transmission to the datacenter can be done simultaneously. Moreover, the data can be saved in multiple locations in parallel. Moreover, with data storage in the datacenter, other operations can also be performed. 6G provides also provides security during data transmission.

\section{Conclusion}
\label{con}
In this article, we have discussed all potential applications of 6G communication technology, and we envision and identify the potential of 6G communication technology in many fields, for instance, transportation, city. It has the potential to revolutionize many technologies and applications. We also explore the impact of 6G communication technology in diverse applications. There are tremendous applications that will fully depend on 6G communication technology, particularly, vehicular technology, healthcare sectors, modern cities and industries. Many new technologies and applications are yet to be conceived due to lack of 6G communication technology. Moreover, we will enter into an intelligent era from smart era and this transition will change our perception on lifestyle, society and business. Furthermore, 6G communication technology will also impact the global economy. Therefore, we will evidence that 6G communication technology will be a game changer technology.

\bibliographystyle{IEEEtran}
\bibliography{mybib}
\begin{IEEEbiography}[{\includegraphics[width=1in,height=1.25in,clip,keepaspectratio]{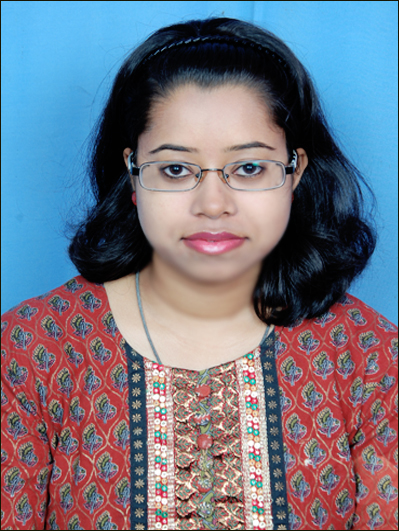}}]{Sabuzima Nayak}(sabuzimanayak@gmail.com) is a research scholar in the Department of Computer Science \& Engineering, National Institute of Technology Silchar, Assam-788010, India. Her research interest is Bioinformatics, Bloom Filter, Communication and Networking, and Big Data. She has published numerous papers in reputed journal, conferences and books.
\end{IEEEbiography}

\begin{IEEEbiography}[{\includegraphics[width=1in,height=1.25in,clip,keepaspectratio]{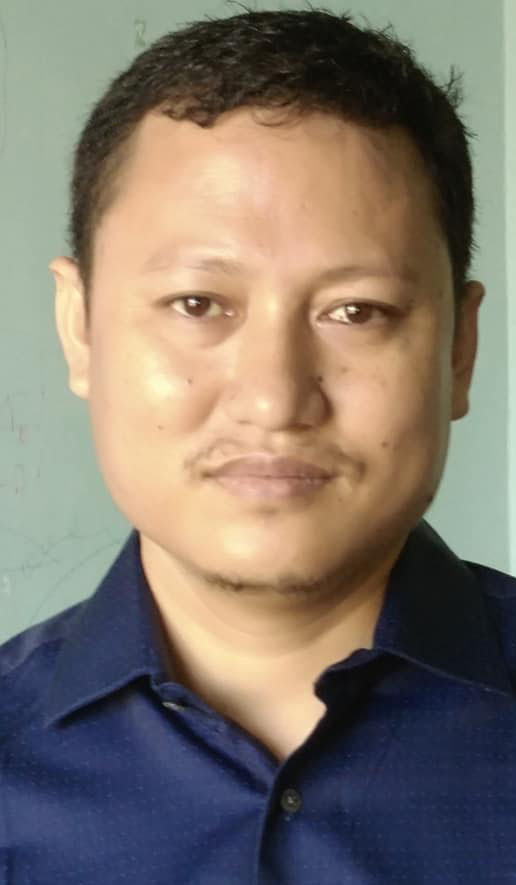}}]{Ripon Patgiri}(ripon@cse.nits.ac.in, patgiri@ieee.org) is currently working as Assistant Professor in the Department of Computer Science \& Engineering, National Institute of Technology Silchar, Assam-788010, India. His research interests are Bloom Filter, Communication and Networks, Big Data, and File System. He has published numerous papers in reputed journal, conferences and books. He is a senior member of IEEE and member of ACM, EAI, and  IETE He has also chaired many conferences.
\end{IEEEbiography}
\vfill
\end{document}